\documentclass[preprint,5p,authoryear]{elsarticle}

\usepackage{graphicx}
\usepackage{natbib} 
\usepackage{amsmath,amssymb}
\usepackage{bbm}
\usepackage{hyperref}
\hypersetup{colorlinks=true,urlcolor=blue,citecolor=blue,pdfborder= 0 0 0}

\bibpunct{(}{)}{;}{a}{}{,}

\def\aj{AJ}%
\def\apj{ApJ}%
\def\apjl{ApJ}%
\def\aap{A\&A}%
\def\mnras{MNRAS}%
\def\jcp{J.~Chem.~Phys.}%
\journal{Astronomy and Computing}

\begin{document}

\begin{frontmatter}

\title{Bisous model -- detecting filamentary patterns in point processes}

\author[a]{Elmo Tempel\corref{cor}}
\ead{elmo.tempel@to.ee}
\author[b,c]{Radu S. Stoica}
\author[a,d]{Rain Kipper}
\author[a,e]{Enn Saar}

\cortext[cor]{Corresponding author}

\address[a]{Tartu Observatory, Observatooriumi~1, 61602 T\~oravere, Estonia}
\address[b]{Universit\'e Lille 1, Laboratoire Paul Painlev\'e, 59655 Villeneuve d'Ascq Cedex, France}
\address[c]{Institut de M\'ecanique C\'eleste et Calcul des Eph\'em\'erides, Observatoire de Paris, 75014 Paris, France}
\address[d]{Institute of Physics, University of Tartu, W. Ostwaldi 1, 51010 Tartu, Estonia}
\address[e]{Estonian Academy of Sciences, Kohtu 6, Tallinn 10130, Estonia}

\begin{abstract}
The cosmic web is a highly complex geometrical pattern, with galaxy clusters at the intersection of filaments and filaments at the intersection of walls. Identifying and describing the filamentary network is not a trivial task due to the overwhelming complexity of the structure, its connectivity and the intrinsic hierarchical nature. To detect and quantify galactic filaments we use the Bisous model, which is a marked point process built to model multi-dimensional patterns.
The Bisous filament finder works directly with the galaxy distribution data and the model intrinsically takes into account the connectivity of the filamentary network. The Bisous model generates the visit map (the probability to find a filament at a given point) together with the filament orientation field. Using these two fields, we can extract filament spines from the data. Together with this paper we publish the computer code for the Bisous model that is made available in GitHub. The Bisous filament finder has been successfully used in several cosmological applications and further development of the model will allow to detect the filamentary network also in photometric redshift surveys, using the full redshift posterior. We also want to encourage the astro-statistical community to use the model and to connect it with all other existing methods for filamentary pattern detection and characterisation.
\end{abstract}


\end{frontmatter}

% \linenumbers

%=============================================================================
\section{Introduction}
\label{sec:intro}

Galaxies, the main building blocks of our Universe, are not uniformly distributed in space. Instead, they form various structures: groups, clusters, chains, filaments, sheets, etc. Galactic filaments are the most prominent part of such a structure, containing nearly half of the total mass of the Universe \citep{Jasche:10,Tempel:14a}.

Until now, the properties of galactic filaments have not yet been utilised fully. Compared, e.g., with galaxy clusters and cosmic voids, filaments are very rarely used as a probe of cosmology and also the role of filaments in galactic evolution is poorly known. In principle, statistics of galaxy filament properties, such as their length, width and connectivity, can be used to measure the large-scale structure and to test cosmological as well as galaxy formation models. However,  detection and definition of filaments has remained problematic so far. Although filamentary structures are easily recognised visually in galaxy survey data, their complicated hierarchical nature does not allow a straightforward mathematical extraction and quantification.

A variety of methods has been proposed (e.g. based on  Minkowski functionals, local topological measures, minimal spanning trees, tessellations, skeleton analysis, kinematics) that attempt to tackle the problem, briefly over\-viewed by \citet{Cautun:14}. These include the methods that classify all the cosmic web elements simultaneously \citep{Hahn:07, AragonCalvo:10, Falck:12, Hoffman:12, Smith:12, Cautun:13, Leclercq:15} or are specifically meant for filament detection \citep{Bond:10, Gonzalez:10, Sousbie:11, Alpaslan:14, Chen:15}.

The first attempts of filament identification have already given some surprising results. For example, filaments have been found in voids \citep{Beygu:13} and other low-density environments \citep[called tendrils:][]{Alpaslan:14}. These examples demonstrate the potential of filament studies to take us closer to understanding the structure formation in the Universe.

In the current paper we use the Bisous model, a probabilistic filament finder that takes an advantage of the Bayesian framework and is straightforward to apply to observational datasets. Our approach to filament detection uses a marked point process that takes into account the connectivity of the filamentary network, i.e. whether or not a given filament is linked to other filament(s). The mathematical basis of the method has been described and proved in \citet{Stoica:05, Stoica:05a, Stoica:07, Stoica:10, Stoica:14}. This model for filament detection has been developed especially for application to observational datasets in cosmology. In \citet{Tempel:14a} we applied the Bisous model to the Sloan Digital Sky Survey (SDSS) data and published a catalogue of filaments for the SDSS\footnote{The catalogue is available at \url{http://cosmodb.to.ee}.}.

The Bisous model fits well in a Bayesian framework that may be considered as an advantage over the conventional methods. The Bisous model does not attempt to classify the web into strict components. Instead, it assigns a confidence estimate to each detected structure. The filamentary network is modelled as a whole and the connectivity between structures is intrinsically implemented in the model. To tackle the large parameter space and global optimisation, the model uses the Markov-chain Monte Carlo (MCMC) sampling together with simulated tempering and simulated annealing.

\citet{Tempel:14c} tested the Bisous model on simulated data. Although they used only the spatial distribution of galaxies/haloes as input, the detected filaments turned out to follow also the underlying velocity field of the simulation, thus indicating that the recovered Bisous filaments are real dynamical structures, not just apparent configurations of galaxies. Similar conclusions were reached  by \citet{Libeskind:15}, using observations of the local Universe.

Filaments are in the non-linear dynamical stage of evolution between the linear and fully virialised objects, and filament evolution in simulations has gained a lot of focus during recent years. For example \citet{Bond:10} analysed the evolution of the distribution of filaments and their properties. They found that most of the filaments are already in place from high redshifts and that most of their evolution is restricted to changes in filament size. \citet{Choi:10} showed that filament widths are most sensitive to the non-linear growth of structure. Recently, an in-depth study of galactic filaments in simulations was made by \citet{Cautun:14}. To move further on, the evolution of actual filaments detected from observations has to be analysed and compared with simulations. This requires advanced observational methods for filament identification, such as the Bisous model.

During recent years, the Bisous model has been extensively used to analyse the filamentary structure in general and to study the influence of the filamentary environment on galaxy/group evolution and formation. \citet{Tempel:13a} and \citet{Tempel:13b} showed that the alignment of major axes of galaxies with respect to galactic filaments depends on galaxy morphology. \citet{Guo:15} showed that isolated galaxies that are located in filaments have up to two times more satellites and the satellites tend to be aligned with galactic filaments \citep{Tempel:15b}. The alignment of structures seems to be an universal trend, having been confirmed in various studies since \citet{Tempel:15a} showed that galaxy pairs in filaments are very well aligned with the underlying structure. The analysis presented in \citet{Tempel:14d} indicates that the distribution of galaxies and groups along the filaments has also some regularity. These successful applications of the Bisous filament finder form a good ground to develop the model further for other astronomical applications. 

The aim of this paper is to review the model presentation, in order to emphasise those mathematical and applied aspects of the Bisous model directly linked with the computational use and numerical implementation of the model. We also want to encourage the astro-statistical community to use the Bisous model, and to compare and connect this model with other methods for filamentary pattern detection and characterisation.

The general outline of the paper is the following. In Sect.~\ref{sec:math} we give a brief description of the mathematical background and tools used. Sect.~\ref{sec:params} explains the motivation and strategies how to choose the parameters for the Bisous model. Sect.~\ref{sec:spines} outlines the algorithm used to extract  single filament spines from the model output. An example is given in Sect.~\ref{sec:example} and the conclusions are presented in Sect.~\ref{sec:con}. 
The computer code for the Bisous filament finder is made available through GitHub\footnote{\url{https://github.com/etempel/bisous}.} and \ref{app:program} gives a brief description how to download and install the program.

%=============================================================================
\section{Mathematical tools}
\label{sec:math}

In this section we briefly describe the main tools we use to detect the filamentary pattern in the galaxy distribution. We outline the key points that are important to understand the code for the Bisous model. The description follows \citet{Tempel:14a}, for details of the mathematical model we refer to \citet{Stoica:05,Stoica:07,Stoica:10}.

%=============================================================================
\subsection{Bisous model}
\label{sect:bisousmodel}

The marked point process we use for filament detection is different from conventional point processes used in the field. The Bisous process models the structure outlined by galaxy positions, not the distribution of galaxies themselves.

We designate $K$ as a finite volume ($0 < \nu(K) < \infty$), where a finite number of galaxies ($\mathbf{d} = \{d_1, \ldots, d_n\}$) are observed. Our aim is to model the filamentary network outlined by the positions of galaxies.

The main hypothesis of our work is that the filamentary network can be modelled by a random configuration of connected and aligned cylinders -- a realisation of a marked point process. Here the points (objects) are the centres of cylinders and marks are the length and orientation of cylinders (given with an uniform law). Note that this is different from the common use of point processes in cosmological studies, where the points are centres of galaxies. In the Bisous model, the centres of galaxies are just used to calculate the probability for filaments (see below).

\begin{figure}
	\includegraphics[width=88mm]{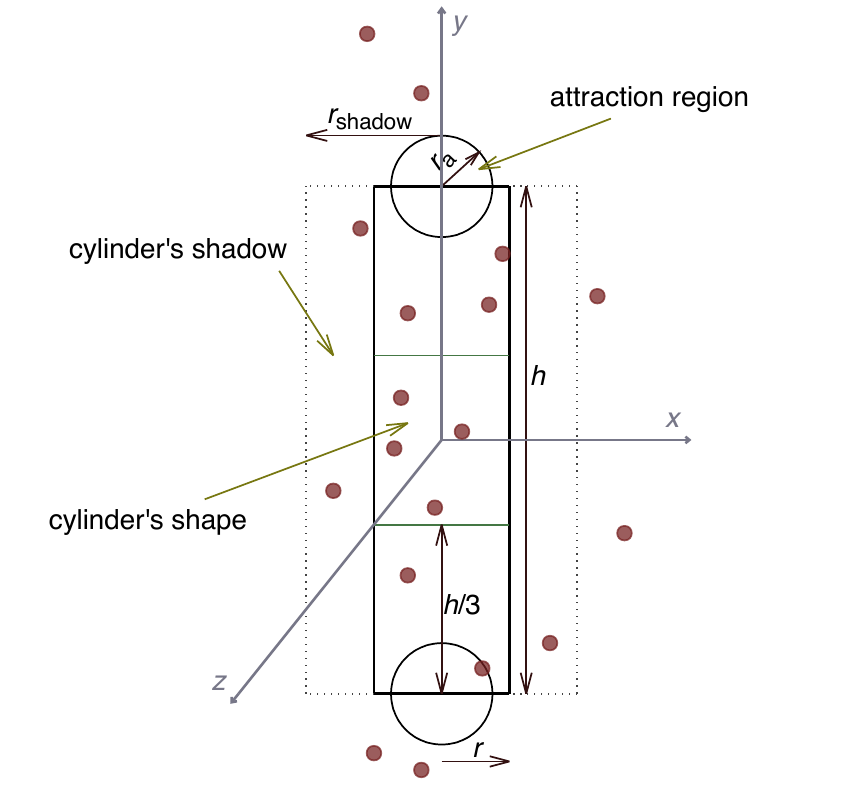}
    \caption{A two-dimensional projection of a cylinder (solid rectangle) with its shadow (dashed lines) within a pattern of galaxies (points). The attraction regions are shown as spheres. The exact shape of the cylinder, its shadow, and the attraction regions depend on the model parameters.}
     \label{fig:cylinder}
\end{figure}

A cylinder is an object given by its centre $k \in K$ and the shape parameters. The shape of a cylinder is characterised by its radius $r$, the length $h$, and the orientation vector $\omega = \left(\sqrt{1-\tau^2}\cos(\eta),\,\sqrt{1-\tau^2}\sin(\eta),\,\tau\right)$.
We denote the cylinder together with its mark (the set of parameters) by $s(y)=s(k,r,h,\omega) \subset K$.

Each cylinder $s(y)$ has two end points. Around these points spheres of radius $r_\mathrm{a}$ are centred, forming the attraction regions. These regions are used to define the connectivity and alignment rules for the model (see Sect.~\ref{sect:intenergy}). The basic cylinder within a field of galaxies is illustrated in Fig.~\ref{fig:cylinder}.

Let $\mathbf{y}=\{y_1=(k_1,m_1),\ldots,y_n=(k_n,m_n)\}$ be a configuration of cylinders, where the cylinder mark is denoted by $m_i$. The ``simplest'' random configuration of cylinders is the stationary Poisson process of unit intensity. This process is constructed in two steps. First, the number of cylinders $n$ is chosen according to a Poisson distribution of the parameter $\nu(K)$. Then the cylinder marks (lengths and the orientation vectors) are chosen, independently and identically distributed with $\nu(M)$, the given marks distribution over the marks space $M$ \citep[see][]{Tempel:14a}. In order to obtain a filamentary network composed of connected and aligned cylinders, we define the probability density:
\begin{equation}
	p(\mathbf{y}|\theta) = \alpha\exp\left[-U(\mathbf{y}|\theta)\right]
	\label{eq:probability}
\end{equation}
where $\theta$ is the vector of parameters, $\alpha$ is the normalising constant, and $U(\mathbf{y}|\theta)$ is the energy function of the system.

The model assumes that locally galaxies may be grouped together inside rather small cylinders that connect and align. Following these ideas the energy function in~(\ref{eq:probability}) can be specified:
\begin{equation}
	U(\mathbf{y}|\theta) = U_\mathrm{i}(\mathbf{y}|\theta) + U_\mathrm{d}(\mathbf{y}|\theta),
\end{equation}
where $U_\mathrm{i}(\mathbf{y}|\theta)$ is the interaction energy controlling the alignments and connections (see Sect.~\ref{sect:intenergy}) and $U_\mathrm{d}(\mathbf{y}|\theta)$ is the data energy controlling the positioning of the cylinders in the galaxy field (see Sect.~\ref{sect:dataenergy}).

In order to specify the model, we have to choose its parameters. Here, the Bayesian
framework is adopted, where the prior for the parameters is denoted by $p(\theta)$
\citep{Stoica:07a,Stoica:07,Stoica:10}. With all these elements, we construct the joint probability density $p(\mathbf{y},\theta)$ and we propose for the filamentary pattern estimator the cylinder configuration maximising it.

%=============================================================================
\subsection{Interaction energy}
\label{sect:intenergy}

The interaction energy term depends on the relative position of the cylinders forming the network and it can be expressed (in general form) as  
\begin{equation} 
	U_\mathrm{i}(\mathbf{y}|\theta)=-n_k(\mathbf{y})\log\gamma_k-\sum\limits_{s=0}^2n_s(\mathbf{y})\log\gamma_s,
	\label{eq:intenergy}
\end{equation}
where $n_k(\mathbf{y})$ is the number of repulsive cylinder pairs and $n_s(\mathbf{y})$ is the number of cylinders connected to the network through $s$ end points. For filament detection, repulsive cylinders are forbidden, hence we may consider $\gamma_k=0$, so for repulsive cylinders it's associated energy is infinite. The variables $\log\gamma_s$ are the potentials associated with the 0-, 1- and 2-connected cylinders.

Two cylinders $y_1$ and $y_2$ ($y_i=(k_i, r_i, h_i, \omega_i)$) are connected if they attract each other, do not reject each other, and are well aligned. Two cylinders attract each other if the distance between the cylinder end points is smaller than the interaction radius $r_\mathrm{a}$ (see Fig.~\ref{fig:cylinder}). Two cylinders are well aligned if $\lvert\omega_1\cdot\omega_2\rvert\ge \tau_{\|}$. We say that two cylinders reject each other if their centres are closer than the minimum allowed distance between cylinders, $d(k_1,k_2)<(h_1+h_2)/2-r_\mathrm{a}$. Two cylinders are considered repulsive, if they are rejecting each other, if they are not orthogonal and if they have roughly the same scale ($1/r_\mathrm{ratio}<r_1/r_2<r_\mathrm{ratio}$), where the ratio of the scales $r_\mathrm{ratio}>1$ is a predefined parameter. Two cylinders are considered to be orthogonal if $\lvert\omega_1\cdot\omega_2\rvert\le \tau_\perp$. The constants $\tau_{\|}\in(0,1)$ and $\tau_\perp\in(0,1)$ are predefined parameters to allow a certain range of mutual angles for aligned and perpendicular cylinders.
See fig.~2 in \citet{Tempel:14a} for illustrations of these definitions.

%=============================================================================
\subsection{Data energy}
\label{sect:dataenergy}

The data energy term describes the local properties of a filament. For that, we use the positions of galaxies to consider whether a cylinder is a fragment of a filament or not. We use several criteria that should be fulfilled simultaneously. The first one is that galaxies should be distributed roughly uniformly along the main axis of the cylinder. The second one is that inside the cylinder the number density of galaxies should be higher than in the close-by neighbourhood (in the shadow) of the cylinder. The third one is that the galaxies belonging to the filament should be concentrated along the cylinder's axis.

In the model, the data energy of a configuration of cylinders  $\mathbf{y}$ is defined as the sum of the energy of each cylinder:
\begin{equation}
	U_\mathrm{d}(\mathbf{y}|\theta) = -\sum\limits_{y\in\mathbf{y}} v(y),
	\label{eq:dataenergy}
\end{equation}
where the potential function of the cylinder $y$ is denoted by $v(y)$.

In order to formulate these requirements we use a test based on counts of galaxies. In short, the score for the test is defined as
\begin{equation}
	p_\mathrm{hyp} = p_\mathrm{u}(y)\cdot p_\mathrm{h}(y) ,
\end{equation}
where $p_\mathrm{u}$ gives the $p$-value for the ``local uniform spread'' and $p_\mathrm{h}$ gives the $p$-value for the ``locally high density''. Both these two tests depend on two more parameters, called the threshold densities $\rho_\mathrm{u}$ and $\rho_\mathrm{h}$, respectively. All the details of the tests are given in \citet{Tempel:14a} and \citet{Stoica:14}.

Additionally, to take into account the spatial distribution of galaxies in a filament we introduce the cylinder concentration
\begin{equation}
\sigma^{2}=\frac{1}{n-2}\sum_{j=1}^n \frac{\delta_j^2}{r^2},
\label{eq:cylinderVariance}
\end{equation}
where $n$ is the number of galaxies in a cylinder, $\delta_j$ is the distance to cylinder's main symmetry axis for the $j$th galaxy, and $r$ is the radius of the cylinder.

Finally, the cylinder potential function $v(y)$ puts together all these criteria  
\begin{equation}
v(y) = 
\begin{cases}
c_\mathrm{hyp}\log \left[p_\mathrm{hyp}(y)\right] - c_\mathrm{con}\sigma^{2}(y) & \text{if} \quad n \geq n_\mathrm{min} \\
-\infty & \text{if} \quad n < n_\mathrm{min}
\end{cases}
\label{eq:datapotential}
\end{equation}
where $n_\mathrm{min} \geq 3$ is a given threshold value, and the parameters $c_\mathrm{hyp}\geq 0$ and $c_\mathrm{con}\geq 0$ are introduced to make the location and the concentration tests numerically comparable. Clearly, the higher $v(y)$ is, the ``better'' is the place to put the cylinder.

%=============================================================================
\subsection{Simulation of the model}
\label{sect:simulation}

To search for the filaments in the galaxy distribution, we need to sample from the joint probability density $p(\mathbf{y},\theta)$. In this purpose we are using an iterative MCMC algorithm, where one iteration consists of two steps. First, we choose a value for the parameter $\theta$, from $p(\theta)$. Second, conditionally on $\theta$, we use the  Metropolis-Hastings (MH) algorithm to sample a cylinder pattern from $p(\mathbf{y}|\theta)$ \citep{Geyer:94, Geyer:99}.

The MH algorithm used is constructed using three types of moves \citep{Lieshout:03, Stoica:05, Stoica:07, Stoica:10}: birth (proposing with a probability $p_\mathrm{b}$ to add a new cylinder), death (proposing with a probability $p_\mathrm{d}$ to delete an existing cylinder), change (proposing with probability $p_\mathrm{c}$ to modify the parameters of an existing cylinder). All the practical details concerning the implementation of the MH dynamics are given in \citet{Tempel:14a} and \citet{Stoica:14}; for a complete description we refer to \citet{Lieshout:03} and \citet{Stoica:05}.

%==================================================================
\subsection{Annealing and tempering}
\label{sect:optimisation}

The MH sampling mechanism is integrated into simulated annealing and simulated tempering algorithms.

Simulated annealing is a global optimisation method. Simulated annealing iteratively samples from $p_n(\mathbf{y},\theta) \propto [p(\mathbf{y}|\theta)p(\theta)]^{1/T_n}$, while the temperature $T_n$ slowly decreases toward zero. For marked point processes, the convergence of a simulated annealing algorithm for a logarithmic cooling schedule was proven by \citet{Stoica:05}. In this algorithm, the  temperature is lowered as $T_n={T_0}/(\log n+1)$, where $T_0$ is the initial temperature and $T_n$ is the temperature at the iteration $n$.

Simulated tempering \citep{Marinari:92} is a multi-temperature simulation technique, where the temperature is changed along a fixed temperature ladder. The temperatures are chosen from the interval $[T_\mathrm{min},T_\mathrm{max}]$, so that $T_\mathrm{n+1}/T_\mathrm{n}=\mathrm{const}$. In theory, simulated tempering requires previous knowledge about the system energies at each temperature, which can be computed during a trial simulation. Another approach is to compute the weights for temperature changes on-the-fly. We utilise the last approach and determine the weights for simulated tempering as described in \citet{Nguyen:13}.

Here we combine both schemes. Since simulated tempering has rather good mixing properties, we use simulated tempering before the simulated annealing stage (see Sect.~\ref{sec:spines}). This strategy helps us to save computational time, since simulated annealing requires for convergence the algorithm initialisation at very high temperatures.

%==================================================================
\section{Choosing the parameters for the Bisous model}
\label{sec:params}

\begin{table*}
	\centering
 	\caption{Parameters that affect the code and should be chosen based on the data and problem at hand. The suggested value (range of values) for each parameter is given in the last column. See text for more details how to choose the optimal parameter values. The list of all parameters is given in the  manual. Parameter names are as used in the configuration file (see \ref{app:program}).}
	\label{tab:params}
	\smallskip
	\begin{tabular}{llc}
		\hline
		Parameter & Short description & Value \\
	    \hline
		\hline
		\multicolumn{3}{c}{\textbf{MCMC parameters}} \\
	    nr\_cycles & Number of MCMC cycles. Temperature is adjusted after every cycle. & $10\,000$ \\
		nr\_moves  & Number of moves in one cycle. Temperature is fixed during one cycle. & $10\,000$$^{a}$ \\
		every\_cycle\_to\_output &  Cylinder configuration is extracted after every $n$th cycle. & $1000$ \\
		cooling\_schedule & Temperature cooling schedule for MCMC. & $[1,2,3,4]$$^{b}$ \\
		temp\_initial & Initial (maximum) temperature. & $2.0\dots 5.0$$^{c}$ \\
		temp\_final & Final (minimum) temperature. & $0.3\dots 1.0$$^{c}$ \\
		prob\_birth & Proposal probability for birth moves in MH. & $0.5$ \\
		prob\_death & Proposal probability for death moves in MH. & $0.3$ \\
		prob\_change & Proposal probability for change moves in MH. & $0.2$ \\
		prob\_birth\_connected & Proposal probability for connected birth for a birth move. & $0.8$ \\
		change\_delta\_r & Maximum shift of the cylinder centre for a change move. & $0.3\dots 0.8$$^{d}$ \\
		change\_delta\_cosi & Max. cosine between the old and new orientations for a change move. & $0.95$ \\
		\multicolumn{3}{c}{\textbf{Data term parameters}} \\
		min\_pts & Number of minimum points inside a cylinder. & $3\dots 5$ \\
		cyl\_rad\_min & Minimum radius of a cylinder (in physical units). & $0.3\dots 0.5$$^{e}$ \\
		cyl\_rad\_max & Maximum radius of a cylinder (in physical units). & $0.5\dots 1.5$$^{e}$ \\
		cyl\_len\_min & Minimum length of a cylinder (in physical units). & $3.0$$^{e}$ \\
		cyl\_len\_max & Maximum length of a cylinder (in physical units). & $10.0$$^{e}$ \\
		cyl\_shadow\_rad & Size of the cylinder shadow region in the units of cylinder radius. & $1.0$ \\
		hyptest\_uniform\_den & Assumed uniform density for hypothesis testing. & $3.0\dots 6.0$$^{f}$ \\
		hyptest\_local\_den & Assumed local density contrast for hypothesis testing. & $3.0\dots 6.0$$^{f}$ \\
		variance\_coeff & Variance coefficient for tje data term. & $0.5\dots 1.0$$^{f}$ \\
		hypothesis\_coeff & Hypothesis coefficient for the data term. & $0.5\dots 1.0$$^{f}$ \\
		\multicolumn{3}{c}{\textbf{Interaction term parameters}} \\
		rad\_connection & Connection radius for cylinders. & $0.5$ \\
		cos\_orthogonal & Maximum cosine for orthogonal cylinders. & $0.5$ \\
		cos\_parallel & Minimum cosine for parallel cylinders. & $0.85$ \\
		lg\_gamma\_0/1/2 & Interaction energy values for 0/1/2-connected cylinders. & $^{g}$ \\
		\multicolumn{3}{c}{\textbf{Spine extraction parameters}} \\
		minimim\_visitmap\_value &  Minimum visit map value for filament spines & $0.05\dots 0.3$ \\
		minimum\_orientation\_strength & Minimum orientation strength for spine detection. & $0.6\dots 0.8$ \\
		\hline
	\end{tabular}
	\begin{list}{}{}
	\item[$^a$] The number of moves in one cycle depends largely on the dataset used. The number of moves should be larger than the actual number of cylinders in one configuration.
	\item[$^{b}$] The available options are: 1~-- constant temperature; 2~-- simulated annealing; 3~-- simulated tempering; 4~-- combined simulated tempering plus simulated annealing.
	\item[$^{c}$] The simulating temperature affects the number of 0-, 1-, 2-connected cylinders in a configuration. These values should be adjusted based on a test simulation.
	\item[$^{d}$] The maximum cylinder shift is given in the units of connection radius. It should be smaller than the connection radius.
	\item[$^{e}$] The cylinder radius determines roughly the  scale of filaments that should be detected. The cylinder radius and size should be chosen so that the cylindrical shape of the cylinder is maintained.
	\item[$^{f}$] These values affect the balance between the different components of the data term. These should be chosen based on the data potential distributions (see Fig.~\ref{fig:datapot}).
	\item[$^{g}$] The interaction energy values affect most severely the balance between the 0-, 1-, 2-connected cylinders. These values should be chosen based on a test simulation since they depend on the dataset used.  Usually,  lg\_gamma\_0 should be positive, lg\_gamma\_1 be around zero, and lg\_gamma\_2 be positive.
	\end{list}
\end{table*}

All the parameters that are needed in order to run the code are given in the configuration file and are explained in the manual provided with the program. Table~\ref{tab:params} shows the main parameters that affect the detection of filaments together with their suggested values. In this section, we give hints for the motivation and strategies, how the parameters that influence the detected filamentary network may be chosen.\footnote{Note that choosing an optimal parameter values is an open mathematical problem.}

\subsection{Parameters for the data potential}
\label{sec:par_data}

The data energy is defined using three quantities: $p_\mathrm{u}$ to describe the local uniform spread of galaxies along the filament, $p_\mathrm{h}$ to describe the locally high density condition, and $\sigma^2$ to describe the cylinder concentration. The values of $p_\mathrm{u}$ and $p_\mathrm{h}$ depend on the threshold densities $\rho_\mathrm{u}$ and $\rho_\mathrm{h}$, respectively. In fact, $\rho_\mathrm{u}$ controls the level of the degree of uniformity inside cylinders, while $\rho_\mathrm{h}$ controls the difference between the intensity inside the cylinder compared with the environs of the cylinder.

To give the hypothesis testing term ($\log p_\mathrm{hyp}$) and the concentration term ($\sigma^2$) equivalent weights, we introduced the constants $c_\mathrm{hyp}$ and $c_\mathrm{con}$. The Fig.~\ref{fig:datapot} shows the influence of these components on the data energy. 

Clearly the data potential depends on the radius and length of the cylinder. These parameters can be fixed or pre-chosen: they can be modelled in a Bayesian framework or estimated based on the galaxy field. The cylinder length is controlled by the mark distribution. For the radius, we determine it following the gradient (perpendicular to the cylinder) of the galaxy density. The radius, where the density gradient is the highest, is taken as the cylinder radius.\footnote{This computation of the radius may slightly change the definition of our mathematical model, but nothing is changed in the philosophy behind it. Instead of a process of cylinders, we now have a process of segments, which we use to build cylinders.}

\subsection{Parameters for the interaction potential}
\label{sec:par_int}

The interaction potential determines the connectivity of the detected filamentary network. The interaction potentials are defined so that connected cylinders are encouraged, while isolated cylinders are rather penalised. This is done by selecting the connection potentials $\log\gamma_s$. These parameters act together with the data term. If they are too strong compared to data energies then we may detect filaments where they do not exist. On the contrary, if connection potentials are too weak, no connected network is formed. Our strategy here is to fix the order of magnitude of the interaction term, based on the data term, and incorporate the interaction term in the Bayesian framework.

\subsection{Parameters for the Metropolis-Hastings dynamics}
\label{sec:mhdyn}

There are several parameters that affect the dynamics of a MH simulation. The quality of the filamentary network detection is defined by the model, but bad dynamics can alter the detection. The proposal probabilities were chosen $0<p_\mathrm{b}, p_\mathrm{d}, p_\mathrm{c} \leq 1$ such that $p_\mathrm{b} + p_\mathrm{d} + p_\mathrm{c}\leq 1$. For simplicity these parameters do not depend on the current state of the sampler. Since we want to detect filaments, we suggest an increased probability for the birth move.

In practice, the number of moves of a MH simulation should be as large as possible (infinite in theory), but, however, it is mainly determined by the time available for computations. Since the MH simulation is complemented by a simulated annealing/tempering optimisation scheme, we also have to determine the number of moves between temperature updates: we suggest that this number should be large enough, at least of the order of magnitude of the number of cylinders in a configuration. 

In the implementation of the MH dynamic, we have to fix the effective sampling volume (the Lebesque measure $\nu(K)$) for the simulation. Since the filamentary structure is rather sparse, it is reasonable to limit the sampling volume to the regions where galaxies exist \citep[see][for more details]{Tempel:14a}.

\subsection{Choosing the temperature values}

Convergence for simulated annealing is guaranteed if the initial temperature is higher than a certain bound that depends on the maximum number of objects in a configuration. We can compute this using hard core repulsion interactions. Nevertheless, this is computationally too expensive. Therefore, we have chosen the temperature values based on several trials and errors. In the code, a set of testing values is proposed.

For simulated tempering, we choose $N_T$ temperature values in the range $T_\mathrm{min}$ and $T_\mathrm{max}$, where the temperature limits are chosen based on a test simulation. For simulated annealing, we set the initial temperature ($T_\mathrm{max}$), where the simulation starts and the final temperature ($T_\mathrm{min}$) at the end of the simulation.

%==================================================================
\section{Statistical analysis of the detected filamentary network}
\label{sec:spines}

The algorithm as described in the previous sections produces $N$ realisations of the filamentary network. In order to improve the quality of statistical inference, the realisations are obtained using both single and parallel MCMC runs. Each realisation includes the information for all cylinders in a configuration. This data is used to construct the statistical tools for the analysis of filamentary network. The proposed tools are the visit and orientation maps, and the algorithm to extract the spines of single filaments.

\subsection{Construction of the visit and orientation maps}

First, the visit map $\mathcal{L}(\mathbf{k})$ estimates the probability that a given point $\mathbf{k}=(x,y,z)$ is touched by the filamentary pattern, defined as
\begin{equation}
	\mathcal{L}(\mathbf{k}) = \frac{1}{N}\sum\limits_{i=1}^N \mathbbm{1}\{\mathbf{k} \in \mathbf{Y}_i\}, 
\end{equation}
where $\mathbf{Y}_1,\ldots, \mathbf{Y}_N$ are $N$ cylinder configurations and $\mathbbm{1}\{\mathbf{k}\in \mathbf{Y}_i\}$ is the indicator function selecting only points that are covered by any of the cylinders in the configuration $\mathbf{Y}_i$. The visit map can be seen as the mean estimate of the filamentary pattern.

Second, for the filamentary network we define the density map $\mathcal{D}(\mathbf{k})$ as a weighted visit map
\begin{eqnarray}
	\label{eq:eq9}
	\mathcal{D}(\mathbf{k}) &=& \frac{1}{N}\sum\limits_{i=1}^N \frac{ \sum\limits_{y \in\mathbf{Y}_i} W(\mathbf{k},y) }
	{\sum\limits_{y\in\mathbf{Y}_i} \mathbbm{1}\{\mathbf{k}\in y\}}, \\
	W(\mathbf{k},y) &=& e^{v(y)} \mathbbm{1}\{\mathbf{k}\in y\},
\end{eqnarray}
where the first summation is over realisations and the next summations are over cylinders in the given configuration $\mathbf{Y}_i$. $W(y)$ defines the weight for a cylinder $y$ in a location $\mathbf{k}$, while the potential function $v(y)$ is defined using Eq.~(\ref{eq:datapotential}).

Third, we define the orientation field $\mathcal{G}(\mathbf{k},\mathbf{\omega})$ for a point $\mathbf{k}$ and for an orientation $\mathbf{\omega}=\phi(\eta,\tau)$:
\begin{equation}
	\mathcal{G}(\mathbf{k},\mathbf{\omega}) =
	\frac{ \sum\limits_{i=1}^N  \sum\limits_{y\in\mathbf{Y}_i}W(\mathbf{k},y) ( \mathbf{\omega}\cdot\omega_y )^2}
	{\sum\limits_{i=1}^N 
	\sum\limits_{y\in\mathbf{Y}_i}W(\mathbf{k},y)},
\end{equation}
where $\mathbf{\omega}\cdot\omega_y$ denotes the scalar product between the cylinder orientation $\omega_y$ and the orientation vector field $\mathbf{\omega}$. The orientation field in a location $\mathbf{k}$ can be calculated for every orientation $\mathbf{\omega}$ and according to the definition $\mathcal{G(\mathbf{k},\mathbf{\omega})}\in [0,1]$. 

Based on the orientation field, we define the density field for orientation strengths. It is defined as the maximum (over $\mathbf{\omega}$) of the orientation field at a given location~$\mathbf{k}$
\begin{equation}
	\mathcal{D_G}(\mathbf{k}) = \max\left\{ \mathcal{G}(\mathbf{k},\mathbf{\omega}) \right\}.
\end{equation}
The corresponding orientation for the maximum value $\mathcal{D_G}(\mathbf{k})$ at the location $\mathbf{k}$ is defined as
\begin{equation}
	\mathbf{\omega}_{\mathcal{G}}(\mathbf{k}) = \arg \max\limits_{\mathbf{\omega}}\left\{ \mathcal{G}(\mathbf{k},\mathbf{\omega}) \right\}.
\end{equation}
Basically, $\mathbf{\omega}_{\mathcal{G}}(\mathbf{k})$ gives the orientation of a filament at the location $\mathbf{k}$ and $\mathcal{D_G}(\mathbf{k})$ measures the strength of the orientation, i.e. whether the orientation is clearly defined or not.

For computing the previous estimator, the orientation $\omega$ is restricted to be one of the orientations of the cylinders in the location $\mathbf{k}$, since the best orientation is always very close to cylinder orientations. This significantly simplifies the estimation of the best orientation in the location~$\mathbf{k}$.

\subsection{Extraction of single filaments}
\label{sec:spine_extraction}

\begin{figure}
	\centering
	\includegraphics[width=88mm]{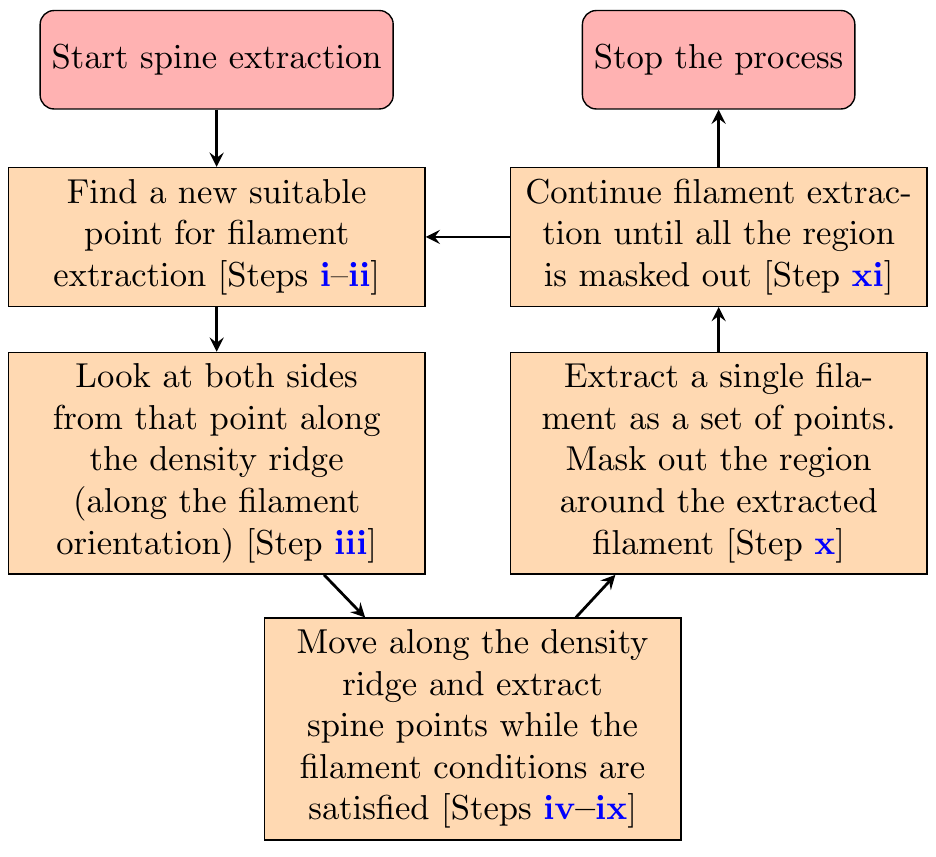}
    \caption{Flow chart that outlines the filament spine extraction algorithm. See Sect.~\ref{sec:spine_extraction} for more information. The roman numbers correspond to the individual steps pointed out in Sect.~\ref{sec:spine_extraction}.}
     \label{fig:spine}
\end{figure}

To extract the spines of single filaments, we use a very similar algorithm to that described in \citet{Tempel:14a}. Basically, this algorithm walks along the ridges given by the filament density map and tests whether the filament orientation is defined and if it coincides with the walking direction. Filaments are extracted as a set of points that form the spines of the filaments. The exact description of the procedure is outlined below and Fig.~\ref{fig:spine} shows the general flow chart for the filament spine extraction algorithm.

For every point $\mathbf{k}$ we have the following values: the filament visit map and the density map values $\mathcal{L}(\mathbf{k})$ and $\mathcal{D}(\mathbf{k})$; the orientation strength $\mathcal{D_G}(\mathbf{k})$ and the filament orientation $\mathbf{\omega}_{\mathcal{G}}(\mathbf{k})$. To extract single filaments from the Bisous model output using these four quantities, we follow the steps as given below.

\begin{enumerate}[i]
	\item We start at a point (designated as $\mathbf{k}_0$) of the highest visit map value $\mathcal{L}(\mathbf{k})$ that is not yet masked out. Initially, all the regions where $\mathcal{L}(\mathbf{k})<\mathcal{L}_\mathrm{lim}$ are masked out, where $\mathcal{L}_\mathrm{lim}$ is the limiting visit map value. The initial visit map is calculated on a grid with a grid step approximately of the scale (radius) of the cylinders. After the maximum is found, the visit map is calculated locally on a finer grid (the final accuracy should be much smaller than the cylinder radius) and an updated maximum in the visit map is found. If the new maximum is farther away from the initial one than half the grid step (the filament spine goes through a nearby grid cell), the point is masked out and a new highest visit map location that is not yet masked out is chosen.
	
	\item We start extracting a filament if the orientation at that point is well defined. The orientation is well defined if $\mathcal{D_G}(\mathbf{k}_0)>\mathcal{D_G}_\mathrm{lim}$, where $\mathcal{D_G}_\mathrm{lim}$ is the limiting orientation strength value. Otherwise, we mask out the region around this point and proceed with the step~(i). The size of the masked region is taken to be the cylinder radius. Masking removes the regions that are already explored.
	
	\item When extracting the filament from the point $\mathbf{k}_0$, we look to both sides along the direction $\mathbf{\omega}_{\mathcal{G}}(\mathbf{k}_0)$.
	
	\item To extract the filament, we move from the point $\mathbf{k}_0$ in the direction $\pm\mathbf{\omega}_{\mathcal{G}}(\mathbf{k}_0)$ by $\delta x$. The new point is designated as $\mathbf{k}_i$. The step size can be arbitrary, but obviously a smaller step produces smoother filaments. The recommended step size $\delta x = r$, where $r$ is the radius of a cylinder.
	 
	\item Next, using Eq.~(\ref{eq:eq9}) we compute the density map $\mathcal{D}(\cdot)$ in the plane that passes through the point $\mathbf{k}_i$ and is perpendicular to the filament direction $\mathbf{\omega}_{\mathcal{G}}(\mathbf{k}_i)$. From that map we find the location of the maximum density (marked as $\mathbf{k}_{i\prime}$) that is closest to the point $\mathbf{k}_i$. This step is necessary to restrict the filament spine to the highest density regions. Additionally, we test whether the point $\mathbf{k}_{i\prime}$ belongs to the masked out region: we stop and proceed with step~(x) if this region has already been explored.
	
	\item We test, whether the orientation is defined ($\mathcal{D_G}(\mathbf{k}_{i\prime})>\mathcal{D_G}_\mathrm{lim}$) at $\mathbf{k}_{i\prime}$. If the orientation is not defined, we stop the algorithm and proceed with the step~(x).
	
	\item If the orientation is defined, we move forward by $\delta x$ and find the next point in the current filament. This new point is used to test two additional criteria.
	
	\item First, to avoid breaks in the filament, we calculate the curvature\footnote{The curvature $\kappa=1/R$, where $R$ is the radius of the sphere that is touching three points.} of the filament at the point $\mathbf{k}_{i\prime}$ using this point and its neighbours. We use the limiting value $\kappa>\kappa_\mathrm{lim}=1/r$ ($r$ is the radius of cylinder) to stop the filament finding algorithm.
	
	\item Second, we require that the orientations at the point $\mathbf{k}_{i\prime}$ and at the neighbouring points $\mathbf{k}_{i\pm 1}$ are similar: $\max\lvert  \mathbf{\omega}_{\mathcal{G}}(\mathbf{k}_{i\prime})\cdot \mathbf{\omega}_{\mathcal{G}}(\mathbf{k}_{i\pm 1}) \rvert >\tau_{\|}$. If these criteria are satisfied, we move in the direction $\pm\mathbf{\omega}_{\mathcal{G}}(\mathbf{k}_{i\prime})$ by $\delta x$ and proceed with the step~(v). Otherwise, we stop the filament finding algorithm and continue with the step~(x).
	
	\item If all the filament points from both sides of $\mathbf{k}_0$ have been found we save the extracted points as a single filament and mask out the region around that filament.
	
	\item We shall return to the first step~(i) again until all the analysed volume is explored (masked out).
\end{enumerate}

\begin{figure}
	\includegraphics[width=88mm]{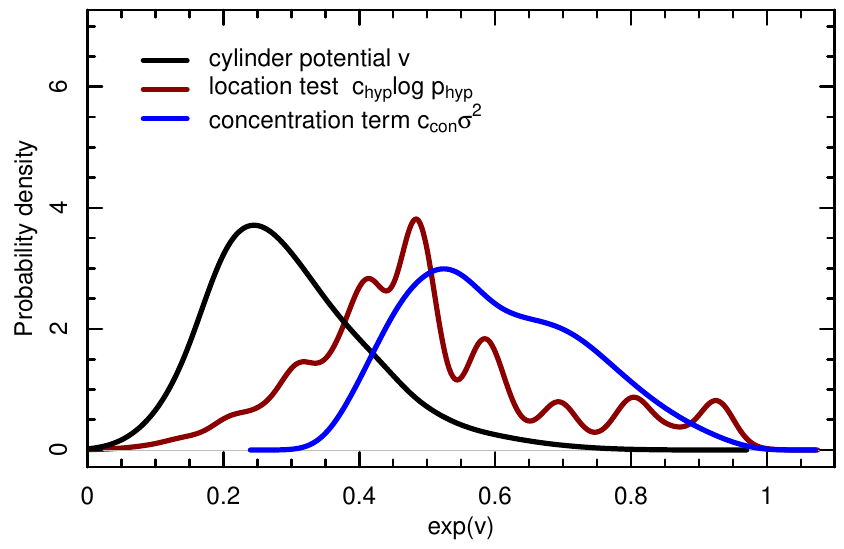}
    \caption{The distribution of the cylinder's potential (black solid line) in the example dataset. The two components -- the location test and the concentration term -- of the cylinder's data energy are shown with red and blue lines, respectively. Note that the curves are smoothed to better show the overall distributions of potentials. Due to the low number of galaxies inside cylinders, the values of the location test potential are discrete and the real distribution has several peaks. The figure illustrates that the location test and the concentration term are equally important in the total potential of the cylinder. The use of the two tests improves the data energy since it is more selective than a single test.}
     \label{fig:datapot}
\end{figure}

\begin{figure}
	\includegraphics[width=88mm]{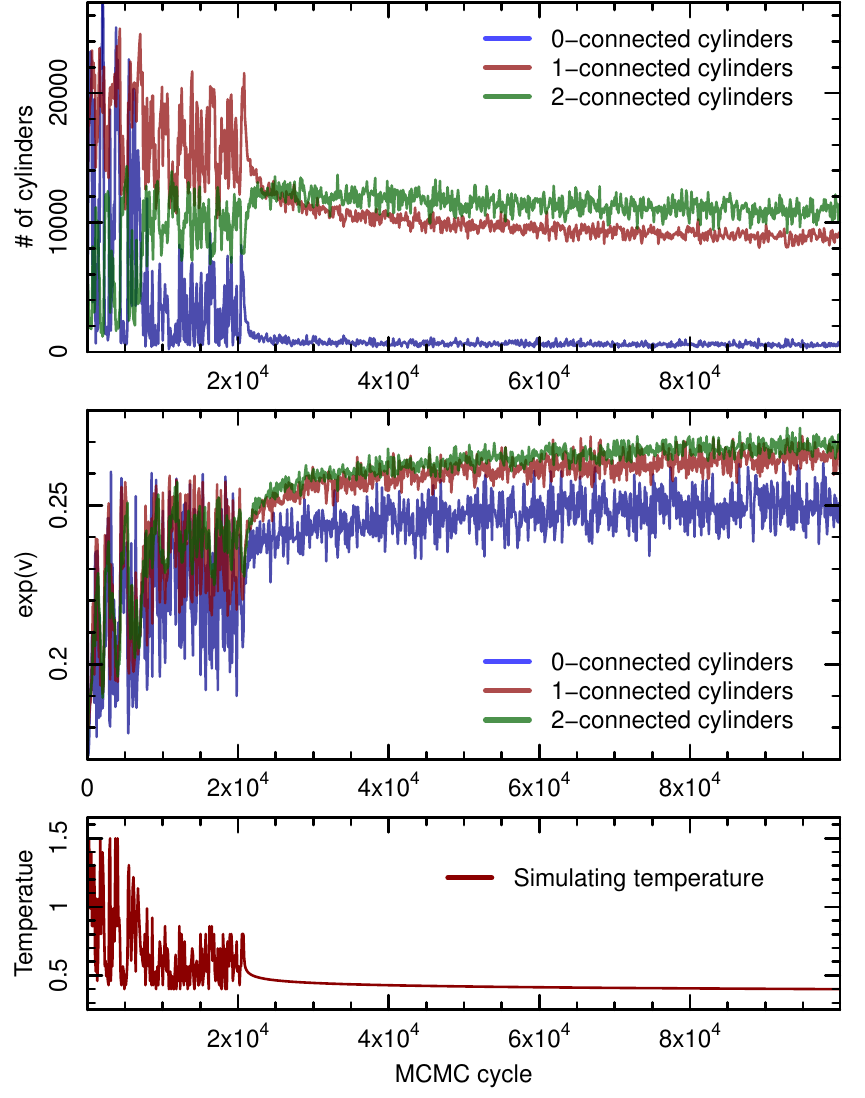}
    \caption{The running statistics of the MCMC simulation. The upper panel shows the numbers of the 0-, 1- and 2-connected cylinders in the model as a function of MCMC cycles. In the middle panel, the average data potentials for the 0-, 1- and 2-connected cylinders are shown. The bottom panel shows the simulated temperature in the model.}
     \label{fig:stats}
\end{figure}

There are two parameters that define the filament spines: $\mathcal{L}_\mathrm{lim}$ defines the limiting visit map density and $\mathcal{D_G}_\mathrm{lim}$ defines the orientation and estimates the strength of orientation for a filament. Both criteria do not affect strong filaments, but they influence the regions where filaments intersect or the regions where filaments are poorly defined. The limiting visit map density should be high enough to have a sufficiently large number of cylinders in a given location for statistical analysis. If the number of realisations is greater than 1000, we suggest to choose $\mathcal{L}_\mathrm{lim}=0.05$, i.e. there should be at least 50 cylinders in every location. The suggested value for the limiting orientation strength  $\mathcal{D_G}_\mathrm{lim}$ is 0.75: this value guarantees that at least half of the cylinders are similarly oriented at a given location.

%==================================================================
\section{Example dataset for the Bisous model}
\label{sec:example}

\begin{figure*}
	\centering
	\includegraphics[width=144mm]{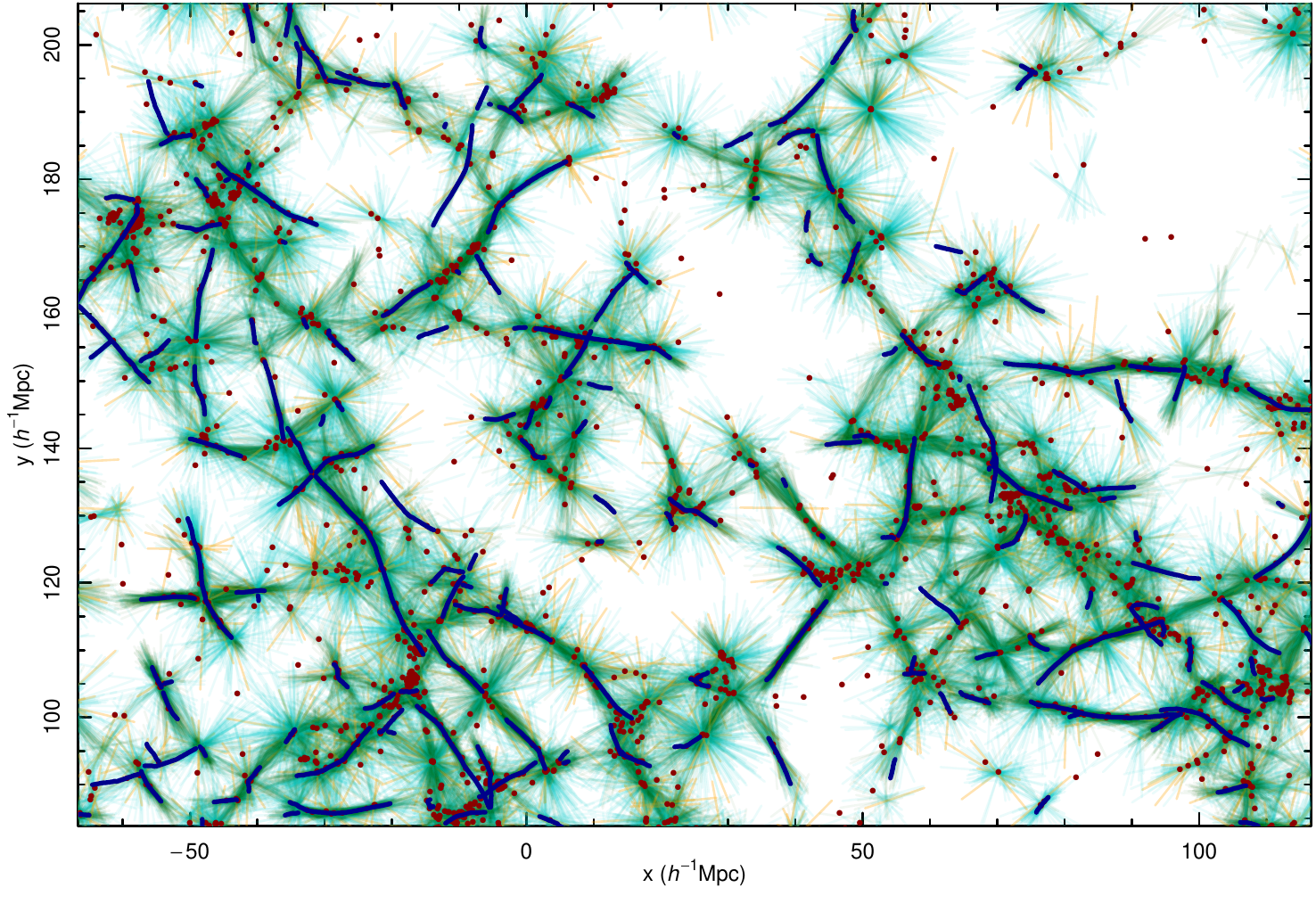}
    \caption{A 5~$h^{-1}\mathrm{Mpc}$ thick slice in the example dataset. To make the filamentary structures  easier to see a smaller area of the full example dataset is shown. Galaxies in this slice are shown as red points. The cylinders in the MCMC realisations are shown as small lines. Altogether there are 500 different realisations. They are colour-coded as following: green for 2-connected, cyan for 1-connected and yellow for isolated cylinders. The extracted filament spines are shown as dark blue lines. The figure shows that the filamentary network is clearly visible, following nicely the distribution of galaxies.}
     \label{fig:slice}
\end{figure*}

The use of the Bisous model is explained using an example dataset accompanied with the program. The example dataset is a 300x320x260~$h^{-1}\mathrm{Mpc}$ brick from the main spectroscopic flux-limited sample of galaxies of the SDSS DR10 down to $m_r=17.77$~mag. For filament detection we suppressed the fingers-of-God redshift distortions (see \citet{Tempel:14b} for the details of the full dataset.) The example dataset includes 179\,463 galaxies. 

The example parameter file\footnote{The example parameter file can be downloaded together with the source code from GitHub (see \ref{app:program}).} is prepared to detect similar filamentary structures as in \citet{Tempel:14a}. The suggested values for the most important parameters are given in Table~\ref{tab:params}.
Since the Bisous model is stochastic, clearly, the realisations are not all identical. Moreover, the practical implementation we propose does not tell us whether we are close to the solution or not. Taking inference from one realisation is not a very efficient strategy. Our results are more robust when using the proposed statistical tools: the visit map, the orientation map, and the filament spines.

We show below some basic statistics and describe the detected filamentary network, using the example dataset. 

Fig.~\ref{fig:datapot} shows the distribution of the data term probabilities together with the location test and the concentration term (see Sect.~\ref{sect:dataenergy}) probabilities for individual cylinders in a detected filamentary network. As mentioned above, the scaling parameters (see Sect.~\ref{sec:par_data}) for the location test and the concentration term should be chosen based on these distributions.

The Bisous model is monitored by following the sufficient statistics\footnote{These statistics can be monitored at runtime since the program constantly updates the statistics file. See the program manual for more details.} -- the numbers of the 0-, 1-, 2-connected cylinders in the configuration. The numbers of these cylinders as a function of the cycles in the MCMC simulation are shown in Fig.~\ref{fig:stats} (upper panel). In the middle panel of Fig.~\ref{fig:stats} we show the average data potentials for the 0-, 1-, 2-connected cylinders. With ``good'' model parameters, the average data potential is the ``best'' for the 2-connected cylinders and the ``worst'' for the 0-connected cylinders. 

Fig.~\ref{fig:stats} also illustrates how simulated tempering and simulated annealing work. At the start of the program, we fixed 20 temperature steps in the range 0.4{\dots}1.5. We can see that the upper temperature value is slightly lowered during the program. If the upper temperature is too high, all the configurations are equally probable, hence, the data and interaction potentials have a weak effect -- we do not detect connected structures. Since the temperature in simulated tempering is jumping up and down, the numbers of the 0-, 1-, 2-connected cylinders also change. At a low temperature, the numbers of the 0-, 1-, 2-connected cylinders do not change dramatically, the configuration is practically frozen.

An example of the detected filamentary network is shown in Fig.~\ref{fig:slice}. In this Figure individual cylinders are shown as small lines and the detected filament spines are highlighted as solid blue lines. We can clearly see that galaxies trace the visit map and the filament spines are aligned with the orientation of cylinders. Fig.~\ref{fig:slice} also shows that some of the galaxies appear to be ``orphans'', not belonging to any cylinder. This happens because the algorithm requires at least three galaxies for a cylinder. Those orphan galaxies are located in  regions, where the number density of galaxies is too low. However, it does not mean that those galaxies do not belong to any filament. They might belong to weaker filaments crossing voids, but are undetected in the Bisous model because of the low number density of galaxies.

\begin{figure}
	\includegraphics[width=88mm]{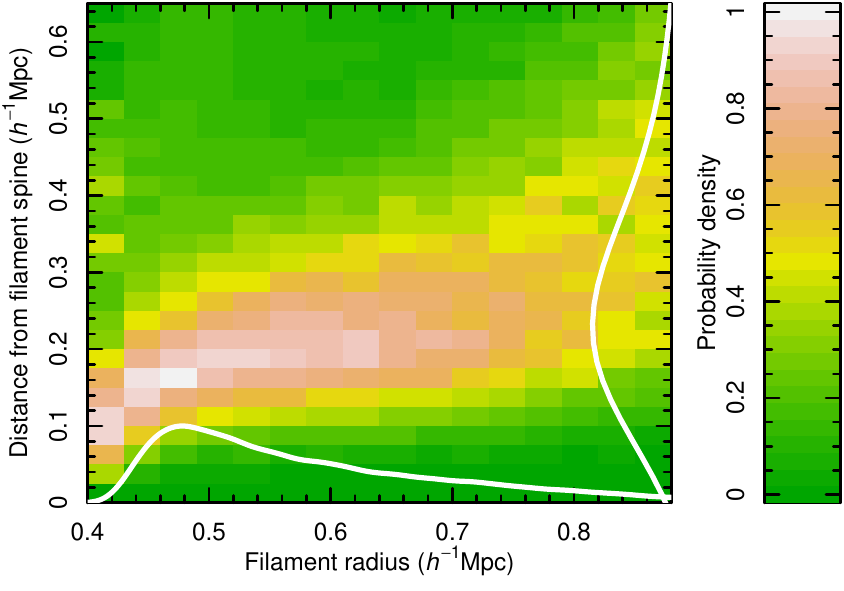}
    \caption{The distribution of galaxy distances from filament spines. The distribution is shown as a function of the radius of the cylinder the galaxy belongs to. The radius of the cylinder roughly defines the filament scale. The marginal distributions are shown as white lines along the axes of the figure. The figure clearly illustrates that galaxies are located close to filament spines.}
     \label{fig:scale}
\end{figure}

Finally, to show how filament spines follow the distribution of galaxies, we calculated the distribution of galaxy distances from the nearest filament spines, as shown in Fig.~\ref{fig:scale}. Since the filament scale (the radius of the cylinder) is not entirely fixed in the model, we show this distribution as a function of the filament radius (the filament scale is defined by the radius of the cylinder). Fig.~\ref{fig:scale} shows that most of the galaxies are located close to filaments and are usually closer than 0.4~$h^{-1}\mathrm{Mpc}$ from the filament spines. The distance of a galaxy from a filament spine is roughly independent of the filament thickness. This is mostly because the variable filament radius in the model helps to detect filaments also in regions where the number density of galaxies is lower. The filament radius in the Bisous model is only a very rough estimate of the true filament thickness.

%==================================================================
\section{Conclusions}
\label{sec:con}

This paper presents the use of the Bisous model, a marked point process with interactions for tracing the filamentary network in the distribution of galaxies. This method works directly on the galaxy distribution and is specifically developed for observational datasets. We review the Bisous model  and emphasise those mathematical and applied aspects of the model, which are directly linked with its computational use and numerical implementation.

Our probabilistic filament finder computes the visit map (the filament detection probability field) and the filament orientation field that
are moments estimators of a complex geometrical pattern. Using these two fields, we define the spines of the filaments and extract single filaments from the data.

In \citet{Tempel:14a} we applied the Bisous model to the SDSS data and showed that the detected filaments fit well with the visible large-scale structure.

Together with the current paper, we publish the computer code of the Bisous model. The code is available in GitHub\footnote{\url{https://github.com/etempel/bisous}.}. \ref{app:program} gives a brief description how to download and install the program.

The current implementation of the Bisous model works with discrete points, i.e. we suppose that the galaxy coordinates are accurate. Since the Bisous model uses the distribution of galaxies to calculate the probability for the filamentary network, the model can be developed further to detect filamentary structure in photometric redshift surveys. In this case, the full posterior for photometric redshifts can be taken into account in the Bisous model. It is highly promising to apply this method to accurate photometric redshift surveys such as J-PAS \citep{Benitez:14}.

\section*{Acknowledgements}

We thank the Referee for a useful feedback that helped to improve the paper. We acknowledge the support by the Estonian Research Council grants IUT26-2, IUT40-2, and TK133.

\appendix

\section{Download and installation details}
\label{app:program}

The computer code is made available through GitHub from the following webpage \url{https://github.com/etempel/bisous}. To download the code, follow the instructions in GitHub.

The program is written in Fortran using the Fortran~2003 standard. The code has been tested using the Intel Fortran Compiler, but it should compile with any Fortran compiler that supports the Fortran~2003 standard. The code can be compiled using the Makefile. If you use Intel Fortran Compiler there is no need to modify the Makefile, otherwise you should specify your Fortran compiler. To compile the program on Unix systems you should just give the command \texttt{make}. This will generate an executable file \texttt{bisous\_program}.

To run the program you should prepare a configuration file, e.g. \texttt{config\_example.ini}. Then just type \texttt{bisous\_program config\_example.ini}.

\section*{References}

%\bibliographystyle{model2-names}
%\bibliography{mybib}

\end{document}